\documentclass[a4paper,12pt]{article}
\usepackage{amssymb,amsmath,amsthm,epsfig,psfrag,amsfonts,mathrsfs}

\newcommand{\RR}{{\mathbb{R}}}

\newcommand{\pa}{\partial}
\newcommand{\dd}{{\rm d}}

\newcommand{\sfrac}[2]{{\textstyle\frac{#1}{#2}}}
\newcommand{\Tr}{\mathrm{Tr}}

\newcommand{\nv}{\mbox{\boldmath{$n$}}}

\newcommand{\R}{{\mathbb{R}}}

\newcommand{\I}{{\mathbb{I}}}

\newcommand{\beq}{\begin{equation}}
\newcommand{\eeq}{\end{equation}}
\newcommand{\bea}{\begin{eqnarray}}
\newcommand{\eea}{\end{eqnarray}}
\newcommand{\ben}{\begin{eqnarray*}}
\newcommand{\een}{\end{eqnarray*}}
\newcommand{\bem}{\begin{enumerate}}
\newcommand{\eem}{\end{enumerate}}

\newcommand{\ra}{\rightarrow}

\newcommand{\cd}{\partial}

\newcommand{\less}{\backslash}

\newcommand{\ph}{{\varphi}}
\def \d{\mathrm{d}}

\newcommand{\ignore}[1]{}

\newcommand{\phvec}{\mbox{\boldmath{$\ph$}}}

\newcommand{\nvec}{\mbox{\boldmath{$n$}}}

\newcommand{\eps}{\varepsilon}

\renewcommand{\div}{{\rm div}\, }
\renewcommand{\ph}{\varphi}
\newcommand{\news}{\setcounter{equation}{0}}

\theoremstyle{plain}

\theoremstyle{definition}

\title{Skyrmions with low binding energies}
\author{Mike Gillard\footnote{email: m.n.gillard@leeds.ac.uk}\hspace*{0.15cm}, 
Derek Harland\footnote{email: d.g.harland@leeds.ac.uk}\hspace*{0.15cm} and 
Martin Speight\footnote{email: speight@maths.leeds.ac.uk} 
  \bigskip
  \\School of Mathematics,
  \\University of Leeds, Leeds LS2 9JT, UK
}
\date{}

\begin{document}

\maketitle

\begin{abstract}
Nuclear binding energies are investigated in two variants of the Skyrme model: the first replaces the usual Skyrme term with a term that is sixth order in derivatives, and the second includes a potential that is quartic in the pion fields.  Solitons in the first model are shown to deviate significantly from ans\"atze previously assumed in the literature.  The binding energies obtained in both models are lower than those obtained from the standard Skyrme model, and those obtained in the second model are close to the experimental values.
\end{abstract}

\section{Introduction}
\label{sec:1}\label{sec:intro}
\news

The Skyrme model is a candidate model of nuclear physics in which nuclei appear as topological solitons.  It can be derived from QCD as an effective theory valid in the limit of an infinite number of colours \cite{witten83b}.  It successfully reproduces a number of qualitative features of nuclei, including quantum numbers of excited states \cite{batmansutwood} and the stability of the alpha-particle \cite{batmansut}.  Quantitative features are generally predicted with reasonable accuracy, but with some notable exceptions.  One class of quantities which are not accurately predicted within the standard Skyrme model is that of nuclear binding energies: typically, these are too large by an order of magnitude.

In the last few years a number of variants of the Skyrme model have been proposed which aspire to rectify this problem.  Among them are a holographic model that incorporates vector mesons \cite{sut-holographic}, the near BPS sextic model \cite{adasanwer}, and the lightly bound model \cite{harland}.  This article contains a thorough investigation of binding energies in the latter two.  These models have in common that they supplement Skyrme's Lagrangian with terms involving the pion fields only.

The Skyrme Lagrangian takes the form
\begin{equation}
\label{Skyrme lagrangian}
 \mathcal{L} = \frac{F_\pi^2}{16\hbar} \Tr(R_\mu R^\mu) + \frac{\hbar}{32g^2} \Tr([R_\mu,R_\nu],[R^\mu,R^\nu]) + \frac{F_\pi^2 m_\pi^2}{8\hbar^3}\Tr(1-U)+\ldots .
\end{equation}
Here $U(x)$ is an $SU(2)$-valued field, $R_\mu=(\pa_\mu U)U^\dagger$ is its right-invariant current, $F_\pi$ and $m_\pi$ denote the pion decay constant and mass, and $g$ is a dimensionless parameter.  This Lagrangian should be understood as defining an effective field theory, and the ellipsis represents additional terms which are higher order either in derivatives or in the pion fields $1-U$.  The baryon number of a field $U(x)$ is defined to be
\begin{equation}
\label{baryon number}
 B(U) = -\frac{1}{24\pi^2}\int_{\RR^3}\epsilon_{ijk}\Tr(R_iR_jR_k) \dd^3 x= 
\frac{1}{2\pi^2}\int_{\RR^3}\mathcal{B} \dd^3 x,
\end{equation}
and is equal to the degree of the restriction of $U$ to any spatial slice.

The sextic model adds to Skyrme's Lagrangian a term sextic in derivatives.  The motivation for doing so is that a Skyrme model with only a sextic term and a potential term is BPS: energies are directly proportional to the baryon number, and hence binding energies are zero.  Thus a Skyrme model that includes a potential and a sextic term with coefficients much larger than the other terms might reasonably be expected to have low binding energies.

The lightly bound model does not include any terms of higher order in derivatives but instead includes an additional potential term proportional to $\Tr(1-U)^4$.  The model whose Lagrangian consists only of this term and the Skyrme term has an exact solution with baryon number 1 but has no solutions of higher baryon number which are stable to fission; nuclei are unbound in this model.  One therefore expects the Skyrme model that includes this potential with a large coefficient to have low binding energies.  Simulations of a two-dimensional toy model support this hypothesis \cite{salsut}.

In the sequel, numerical approximations to energy minima in both models will be presented for baryon numbers one to eight.  These results are consistent with the hypotheses that low binding energies are attainable in both models.  However, the sextic model has proved more challenging to simulate numerically and we have not been able to obtain reliable results for soliton energies in the parameter regime in which binding energies are expected to be comparable to those of
real nuclei.  The structures of energy minima obtained in the sextic model differ radically from those previously assumed in the literature \cite{adasanwer,adasanwer2,adanaysanwer,bonmar,bonharmar}; thus a re-examination of the results obtained in these papers seems to be warranted.  The energy minima obtained in the lightly bound model differ from those in the standard Skyrme model in two respects: their symmetry groups are smaller; and they have the structure of ensembles of particles.  The sextic model will be discussed in section \ref{sec:2} and the lightly bound model in section \ref{sec:3}; some conclusions will be drawn in section \ref{sec:4}.

\section{The sextic model}
\label{sec:2}\label{sec:sextic}
\news

In this section we study a one-parameter family of models with energy
\bea
E_\eps&=&\int_{\R^3}\bigg(-\frac\eps4\Tr R_iR_i+\frac12\left(\frac{1}{12}
\epsilon_{ijk}\Tr(R_iR_jR_k)\right)^2\nonumber \\
&&\qquad\qquad\qquad\qquad\qquad\qquad+\frac18\left(\Tr(1-U)\right)^2\bigg)
\dd^3 x,
\eea
$\eps\geq 0$ being the parameter. So, we have replaced the usual Skyrme term
with a sextic term proportional to $\mathcal{B}^2$, and
modified the potential.
The regime of interest is when $\eps$ is positive and small. The choice of
potential is motivated by two facts:
skyrmions in the $\eps=0$ model are {\em compactons}, which is numerically
convenient (strongly localized solitons are less prone to finite box effects),
and this potential is {\em massless}. That is, the expansion of the potential
about $U=1$ has no quadratic terms, so the pions in this theory are massless.
This is important because, were the pion mass non-zero, it would scale like
$\eps^{-1}$, and hence would be unphysically large in the regime of
interest.

 It is convenient for our purposes to think of the
Skyrme field as taking values in $S^3$ rather than $SU(2)$. Thus we
define $\ph_0,\ldots,\ph_3$ so that
\beq
U=\ph_0\I_2+i\ph_1\tau_1+i\ph_2\tau_2+i\ph_3\tau_3
\eeq
where $\tau_1,\tau_2,\tau_3$ are the Pauli matrices, and $\ph_0^2+\ph_1^2
+\ph_2^2+\ph_3^2=1$. In terms of the field $\ph:\R^3\ra S^3\subset\R^4$, the
energy of interest is
\bea
E_\eps&=&\frac12\int_{\R^3}\left(\eps|\d\ph|^2+|\ph^*\Omega|^2
+{\cal U}(\ph)^2\right)
\dd^3 x,
\label{eq:nearbps}
\eea
where $\Omega$ is the usual volume form on the unit 3-sphere and
\beq\label{ciat}
{\cal U}:S^3\ra \R,\qquad {\cal U}(\ph)=1-\ph_0.
\eeq
This formulation of the model makes manifest the simple, but crucial, fact
that $E_0$ is invariant under 
volume-preserving diffeomorphisms of $\R^3$ \cite{adasanwer}:
given any field
$\ph$ and any diffeomorphism $\psi:\R^3\ra\R^3$ such that
\beq
\psi^*(\dd x_1\wedge\dd x_2\wedge\dd x_3)=\dd x_1\wedge\dd x_2\wedge\dd x_3,
\eeq
$E_0(\ph\circ\psi)=E_0(\ph)$. It will be convenient to introduce the 
shorthand
\beq
E_\eps=\eps E^{(2)}+E^{(6)}+E^{(0)},
\eeq
the superscript indicating the degree of the corresponding piece of the
energy density as a polynomial in spatial partial derivatives.

As is well known \cite{artroctchyan,adasanwer,spe-rhm}, 
in the case $\eps=0$ this model is BPS:
\beq
E_0\geq c_{\cal U}B,\qquad\qquad c_{\cal U}=\int_{S^3}{\cal U},
\eeq
with equality if and only if
\beq
\ph^*\omega=U(\ph)\, \dd x_1\wedge\dd x_2\wedge\dd x_3,
\label{eq:bog}
\eeq
or, equivalently, $\mathcal{B}={\cal U}$.
For our choice of potential, $c_{\cal U}=Vol(S^3)=2\pi^2$ (since $\int_{S^3}\ph_0
=0$) and solutions of (\ref{eq:bog}) have compact support, meaning
$\phi(x)=v=(1,0,0,0)$, the vacuum value, for all $x$ outside some 
bounded subset
of $\R^3$.  
Equation (\ref{eq:bog}) can be interpreted as the condition that
$\ph$ is a volume-preserving map from the subset of $\R^3$ on which 
$\phi\neq v$ to $S^3\less\{v\}$ equipped with the deformed volume form
$\Omega'_{\cal U}=\Omega/{\cal U}$. From 
this we deduce that any $B=1$ BPS skyrmion
(solution of (\ref{eq:bog})) occupies a total volume of
\beq
Vol_1=\int_{S^3\less\{v\}}{\Omega'_{\cal U}}=4\pi^2
\eeq
and $Vol_B=B\, Vol_1$, in obvious notation. 
There is a $B=1$
solution within the hedgehog ansatz,
\beq
\ph_H(r\nvec)=(\cos f(r),\sin f(r)\nvec),
\eeq
where the profile function has $f(0)=\pi$, $f(\infty)=0$, and satsifies the
ODE
\beq
-\frac{df}{dr}\frac{\sin^2 f}{r^2}=1-\cos f.
\eeq
An implicit solution to this boundary value problem is given by
\beq
-\sfrac13 r^3 = f-\pi+\sin f.
\eeq
There are many ways to construct charge $B$ BPS skyrmions.
One can, for example, superpose $B$
charge 1 solutions (e.g.\ hedgehogs) with disjoint support. Less trivially,
one can precompose a charge 1 solution with a volume-preserving
$B$-fold covering map $\R^3\less\R\ra\R^3\less\R$, for example
\beq
\psi_B:\R^3\less\R_z\ra\R^3\less\R_z,\qquad
\psi_B:(r,\theta,\phi)\mapsto(B^{-1/3}r,\theta,B\phi),
\label{eq:yeahright}
\eeq
in spherical polar coordinates. Previous phenomenological
studies of BPS \cite{adasanwer,adasanwer2,adanaysanwer} and near-BPS
\cite{bonmar,bonharmar} skyrmions have all been based on BPS skyrmions
of this type. Note that, by their very construction, these BPS skyrmions 
have spherically symmetric baryon density (they start with a hedgehog, then
precompose with a map which preserves $\mathcal{B}$).

To simulate the model with $\eps>0$
numerically, we put it on a cubic lattice of size
$N^3$ with lattice spacing $h$, typical values being $N\approx151\sqrt[3]{B}$ and
$h=0.05$, 
so that the box length $hN$ is, for each $B$, roughly twice the radius of the
BPS skyrmion given by (\ref{eq:yeahright}). 
Dirichlet boundary conditions are imposed (so $\ph=v$ on the
boundary of the cube).  We replace spatial partial derivatives by 
4th order accurate difference operators to obtain a lattice approximant to
$E_\eps$. Starting at $\eps=1$ with some initial choice of $\ph$, this
lattice energy is then minimized using a gradient-based minimization
algorithm (the quasi-Newton L-BFGS method, limited memory version). Having found a (perhaps only local)
minimum of $E_1$, we then decrease $\eps$ slightly and, using the $E_1$ 
minimizer
as initial data, minimize again. 
Iterating this process we construct a curve of local
$E_\eps$ minimizers, parametrized by $\eps$. To increase our chance of
finding the global minimizer of $E_\eps$, we repeat the process with a
variety of initial guesses at $\eps=1$ 
(generated by, {\em inter alia}, the rational map ansatz, and the
product ansatz applied to sets of lower charge solutions),
obtaining conjectured global minimizers for all charges from 1 to 8.\,
Using the same
differencing scheme, we also construct a lattice approximant $B_{num}$ to
$B$. For all the results we will report we 
find that this is accurate (i.e.\ integer valued) to within
$0.01$ \% but is always lower than $B$. 
We take this as a rough measure of the expected accuracy of our
energies. Finite box simulations generically underestimate energies because they
exclude the contribution of the soliton tail. We compensate for this by
reporting $E=(B/B_{num})E_\eps$.

 As well as monitoring the accuracy of $B$, we also 
keep track of the Derrick scaling constraint, as in \cite{jayspe}. 
That is, any critical point of $E_\eps$ should, by the Derrick scaling
argument \cite{der}, satisfy the virial constraint
\beq
D(\eps):=\frac{1}{E_\eps}(3E^{(6)}(\eps)-\eps E^{(2)}(\eps)-3E^{(0)}(\eps))=0.
\label{eq:virial}
\eeq
In a finite box, large enough that boundary pressure is negligible, $D(\eps)$
should be small and positive \cite{jayspe}. If $D(\eps)$ becomes negative,
or large and positive, this indicates that our numerical solution is not
reliable, and we discard it. In all the simulations reported here, we find that
$D(\eps)$ remains small (less than 1.5\%) for all $\eps\in[0.2,1]$ but then
explodes shortly below $\eps=0.2$ when $B>1$. Simultaneously to the bad
violation of $D(\eps)=0$, we find that the numerical solutions develop
spike-like singularities, indicating the development of spatial structures
smaller than the resolution of our lattice. Once this occurs, we can no longer
track the solution curve. We conclude, then, that the model is numerically
inaccessible (at least with our methods) for $0\leq\eps<0.2$. We will
see shortly that, for $B=1$ and $B=2$, other methods can be used that
allow us to get much closer to $\eps=0$, but these rely on the enhanced
symmetry of these cases.

\begin{figure}[htb]
\begin{center}
\includegraphics[width=13.5cm]{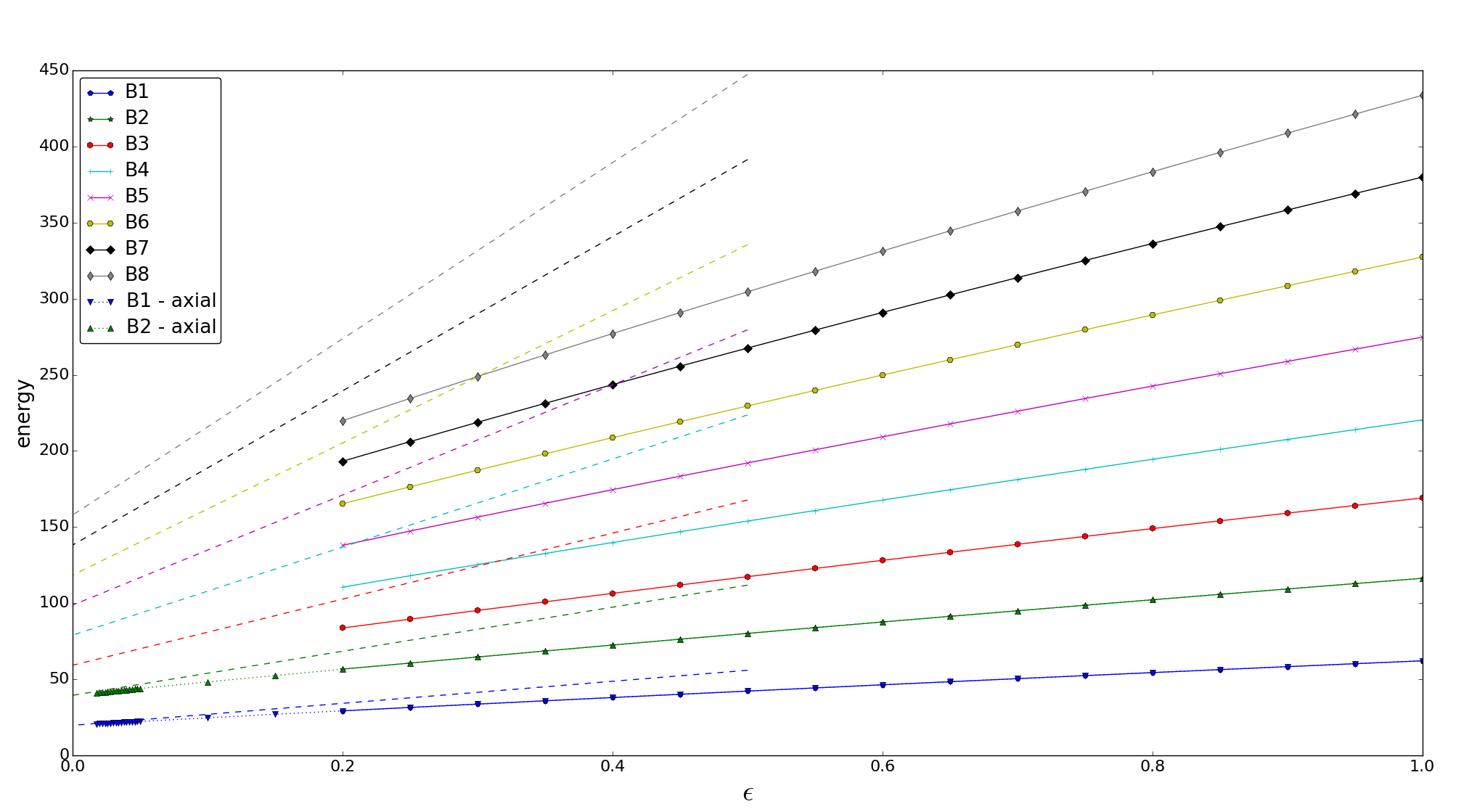}
\end{center}
\caption{Minimum energies in the charge $B$ sector for $B=1,\ldots,8$, as a function of $\eps$. The dashed lines show the upper bound derived in equation (\ref{eq:ub}). The dotted lines show the minimum energies for $B=1,2$ as a function of $\eps$ when restricted to axial symmetry.}
\label{fig:energies}
\end{figure}

Let us denote by $E_{min}(B,\eps)$ the energy of the global minimizer of
$E_\eps$ of charge $B$ (assuming that such exists). It is clear from the 
definition of $E_\eps$ that this is a monotonically increasing function of
$\eps$, and that $E_{min}(B,0)=2\pi^2 B$. It is bounded above by the
total energy of any charge $B$ BPS solution, for example, a superposition
of $B$ hedgehogs of disjoint support. Hence
\beq
E_{min}(B,\eps)\leq B(2\pi^2+\eps E^{(2)}(\ph_H))\approx  (19.74+72.36\eps)B,
\label{eq:ub}
\eeq
so $E_{min}(B,\eps)$ converges to $2\pi^2B$ as $\eps\ra0$.
Figure \ref{fig:energies} shows a plot of $E_{min}(B,\eps)$ for 
$B=1,\ldots,8$, $0.2\leq\eps\leq 1$, as predicted by our numerical
data. The energy value for $B=8$ is taken from the minimum energy configuration at each sampled value of $\eps$. Energy and baryon number data for $\eps=1,0.2$ can be found in appendix \ref{sec:6}, table \ref{tb:bpsdata}.

\begin{figure}[!ht]
\begin{center}
\includegraphics[width=13.5cm]{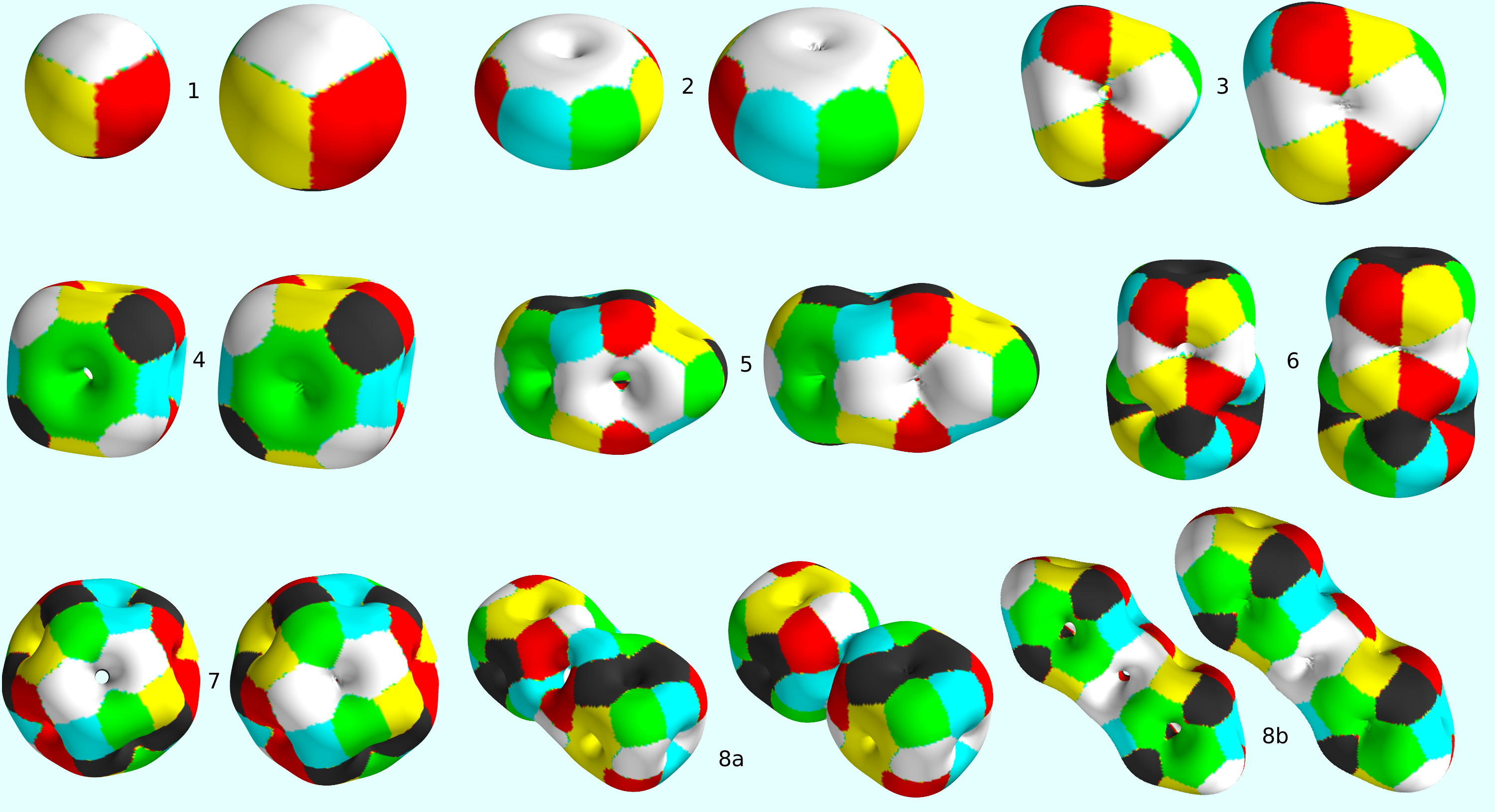}
\end{center}
\caption{Level sets of baryon density for skyrmions in the sextic model with $\eps=1$ (left) and $\eps=0.2$ (right), degrees $B=1$ to $B=8$.
On each surface $\mathcal B = 0.5\max\{\mathcal B(x)\}$. 
}
\label{fig:leaf}
\end{figure}

Figure \ref{fig:leaf} presents, for each degree from 
$B=1$ to $B=8$, a pair of energy minimizers, for $\eps=1$ (left image) and $\eps=0.2$ (right image). In each case, a surface of constant baryon density
is shown, and the colouring of each point represents the orientation
of $\phvec=(\ph_1,\ph_2,\ph_3)$ there. Since the $B=1$ minimizers are
within the hedgehog ansatz, their pictures specify the colouring scheme: we
radially project $\phvec$ onto the unit cube 
$\max\{|\ph_1|,|\ph_2|,|\ph_3|\}=1$ and colour each face of the cube: where $(\pm 1,0,0)$ are yellow/cyan, $(0,\pm 1,0)$ are red/green and $(0,0,\pm 1)$ are white/black. Except for $B=6$, the 
$\eps=1$ minimizers are
qualitatively similar to those of the usual Skyrme model with a pion mass:
in both cases, the potential strongly penalizes fields where $\phi$ is close
to $-v$ over large regions, and hence, disfavours the hollow shell-like
minimizers found in the Skyrme model without a potential, preferring the
formation of chain-like structures. (The $B=6$ skyrmion suggests that this
effect becomes important at lower $B$ for our choice of potential than
for the standard pion mass potential.)
Swapping the quartic Skyrme term
for the sextic term does not seem to have made a qualitative difference to 
the skyrmions. A similar observation was made by Floratos and Piette
\cite{flopie}, who found that the energy minimizers (for $B=1$ to $5$) in the
sextic model with {\em no} potential were qualitatively similar to those
of the usual {\em massless} Skyrme model. We present two local minimizers
for $B=8$, labelled $8a$ and $8b$. At $\eps=1$ it is clear that $8a$ has
lower energy (by 0.35\%), but at $\eps=0.2$ the energy difference is close to
our numerical accuracy, and $8b$ is slightly lower.

\begin{figure}[!ht]
\begin{center}
\includegraphics[scale=0.25]{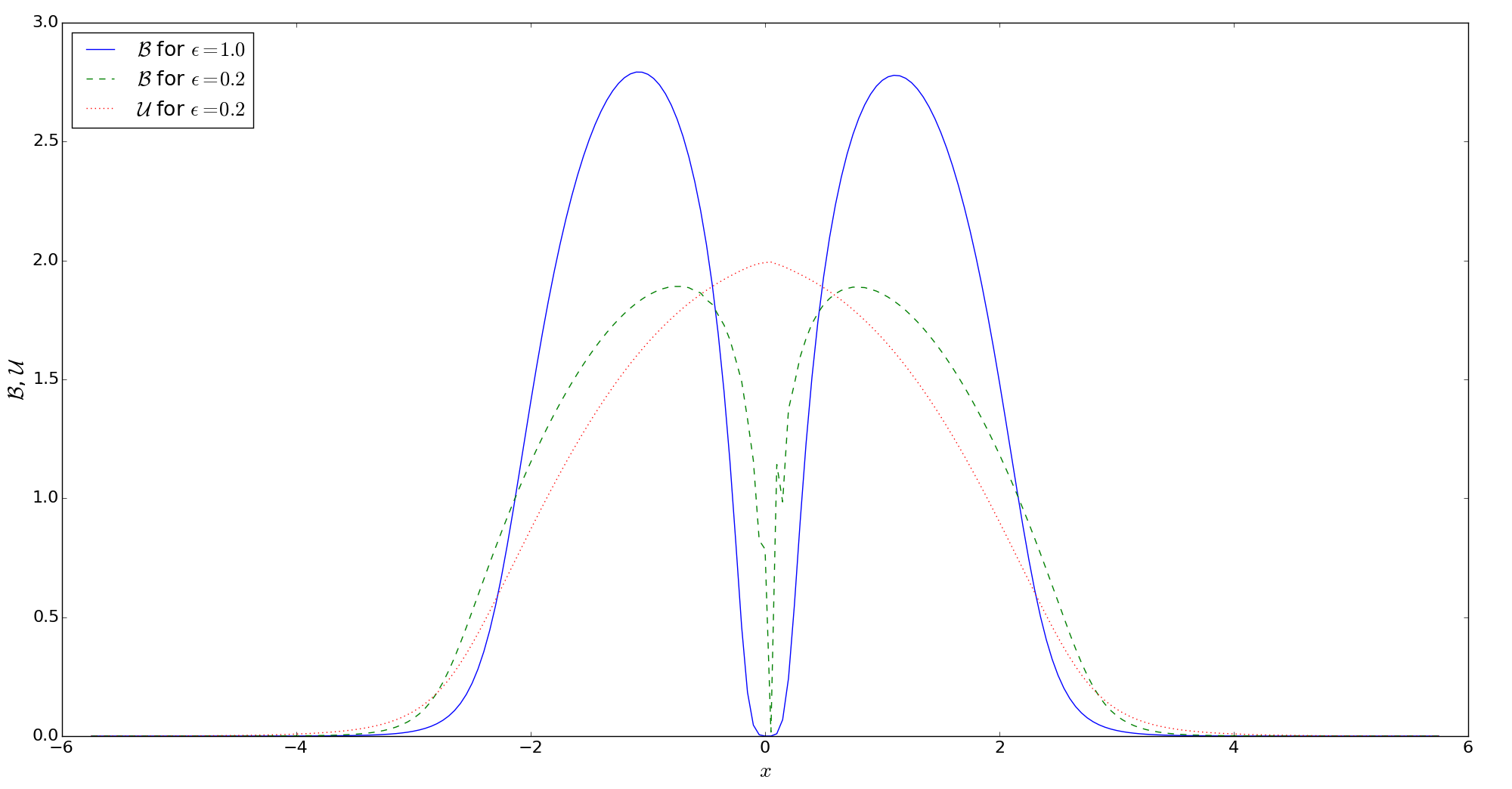}
\end{center}
\caption{Baryon density on a diagonal line through the centre of the $B=4$
skyrmion for $\eps=1$ and $\eps=0.2$. The dotted red curve shows 
${\cal U}(\ph(x))$ for $\eps=0.2$, where ${\cal U}$ is the 
square root of the potential (see equation \ref{ciat}). 
}
\label{fig:agalop}
\end{figure}

Comparing the $\eps=0.2$
minimizers with their $\eps=1$ counterparts, we find that as $\eps$ reduces,
the skyrmions have a tendency to lose symmetry,
the holes in the level sets of $\mathcal{B}$ tend to shrink,
and $\mathcal{B}$ tends to become more uniform within the
skyrmion core. The density of nuclear matter is roughly constant in real 
nuclei, so this effect is desirable, since it reproduces at a classical level a feature that may emerge only after quantisation in the conventional Skyrme 
model. It is along
rays emerging through the holes, on which $\mathcal{B}=0$,
 that singularities begin to form
as $\eps$ drops below $0.2$. This is consistent with the hypothesis that,
as $\eps\ra 0$, the minimizers converge to low regularity solutions of
the Bogomol'nyi equation (\ref{eq:bog}): such a solution must develop
singularities wherever $\mathcal{B}=0$ and $\ph\neq v$. A more detailed
picture of this process can be seen in figure \ref{fig:agalop}, which shows
plots of $\mathcal{B}$ along a line through the centre of the $B=4$ skyrmion,
again for $\eps=1$ and $\eps=0.2$. This line pierces the cube diagonally
through a pair of maximally separated corners. Also plotted is the 
(square root) potential density $\mathcal{U}$ for $\eps=0.2$, suggesting that
$\ph$ is converging, as $\eps\ra0$, to a continuous but not differentiable
solution of $\mathcal{B}=\mathcal{U}$.

\begin{figure}[!ht]
\begin{center}
\includegraphics[width=13cm]{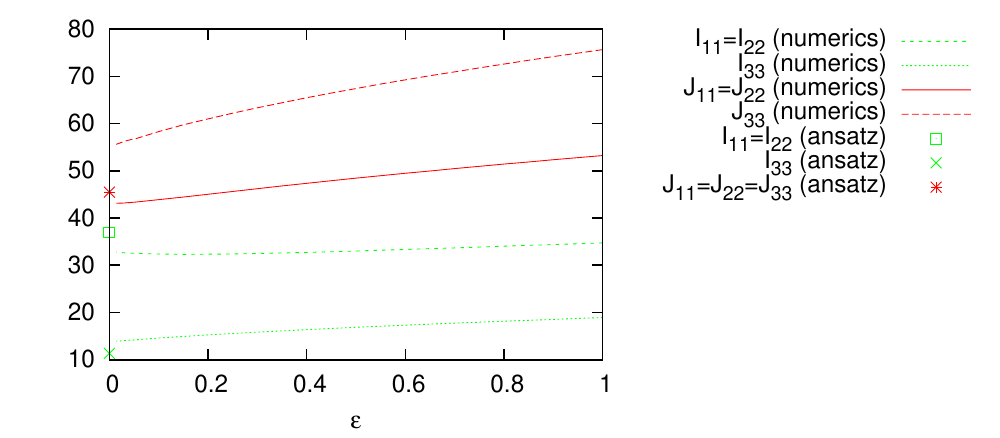}
\end{center}
\caption{Moments of inertia for the skyrmion with $B=2$.  The non-zero components of the leading terms in the $\varepsilon$-expansion given in \eqref{eq:moments} are plotted for a range of values of $\varepsilon$, together with the predictions from the ansatz $\ph_H\circ\psi_2$.}
\label{fig:inertia}
\end{figure}

 Except for $B=1$ and $B=2$, there is no
evidence that $\ph$ is converging to a BPS skyrmion with axial symmetry.
In the case $B=2$, $\ph$ appears to converge to a BPS skyrmion with
axial symmetry, but not one for which $\mathcal{B}$ is spherically
symmetric. This casts doubt on the validity of phenomenological studies of
BPS and near BPS skyrmions based on fields of the form $\ph_H\circ\psi_B$
(see equation (\ref{eq:yeahright}))
\cite{adasanwer,adasanwer2,adanaysanwer,bonmar,bonharmar}. Such studies
perform a rigid body quantization, and depend strongly on
the isospin and spin inertia tensors of the BPS
skyrmions used, 
\bea
I_{ij}(\ph)&=&\int_{\R^3}\d\ph_0\wedge\d\ph_i\wedge *(\d\ph_0\wedge\d\ph_j)+O(\eps),
\nonumber \\
J_{ij}(\ph)&=&\int_{\R^3}{\mathcal{B}(x)}^2(r^2\delta_{ij}-x_ix_j)\dd^3x+O(\eps),
\label{eq:moments}
\eea
respectively. For $\ph_H\circ\psi_B$, $J$ in particular is isotropic
to leading order,
$J=\mbox{const}\, \I_3+O(\eps)$, a very strong constraint which is not supported
by the numerical data: see figure \ref{fig:inertia}.

Since all the minimizers are axially symmetric
for $B=1$ and $B=2$, the $E_\eps$ minimization problem can be
reduced to two dimensions
in these cases. We seek minimizers of the
form
\beq
\ph(\rho,z,\phi)=(n_1(\rho,z),n_2(\rho,z)\cos B\phi, n_2(\rho,z)\sin B\phi,
n_3(\rho,z))
\label{eq:axan}
\eeq
where $(x_1,x_2,x_3)=(\rho\cos\phi,\rho\sin\phi,z)$, and $\nv$ is a map
from the right half-plane to $S^2$ which maps the $z$-axis onto the great circle $n_2=0$. The energy of such a map is
\bea
E_\eps(\nv)&=&2\pi\int_{(0,\infty)\times\R}\bigg\{
\eps(|\cd_\rho\nv|^2+|\cd_z\nv|^2+\frac{B^2}{\rho^2}n_2^2)\nonumber\\
&&\quad+
\frac{B^2}{\rho^2}(\cd_\rho n_1\cd_z n_3-\cd_z n_1\cd_\rho n_3)^2+(1-n_1)^2
\bigg\}\rho\dd\rho\dd z.
\eea
This integral was minimised using a (zero temperature) simulated annealing algorithm.  A logarithmic lattice was used for $\rho$ with lattice spacings $h=0.05$ in $\ln\rho$ and $z$.  The ranges of $\ln\rho$ and $z$ were [-5,2] and [0,7.5], and partial derivatives were replaced by first order differences.  In practice we found that minimizing a discretisation of the difference between the energy and its topological lower bound, rather than of the energy, improved the performance of the algorithm.

\begin{figure}[htb]
\begin{center}
\includegraphics[scale=1.0]{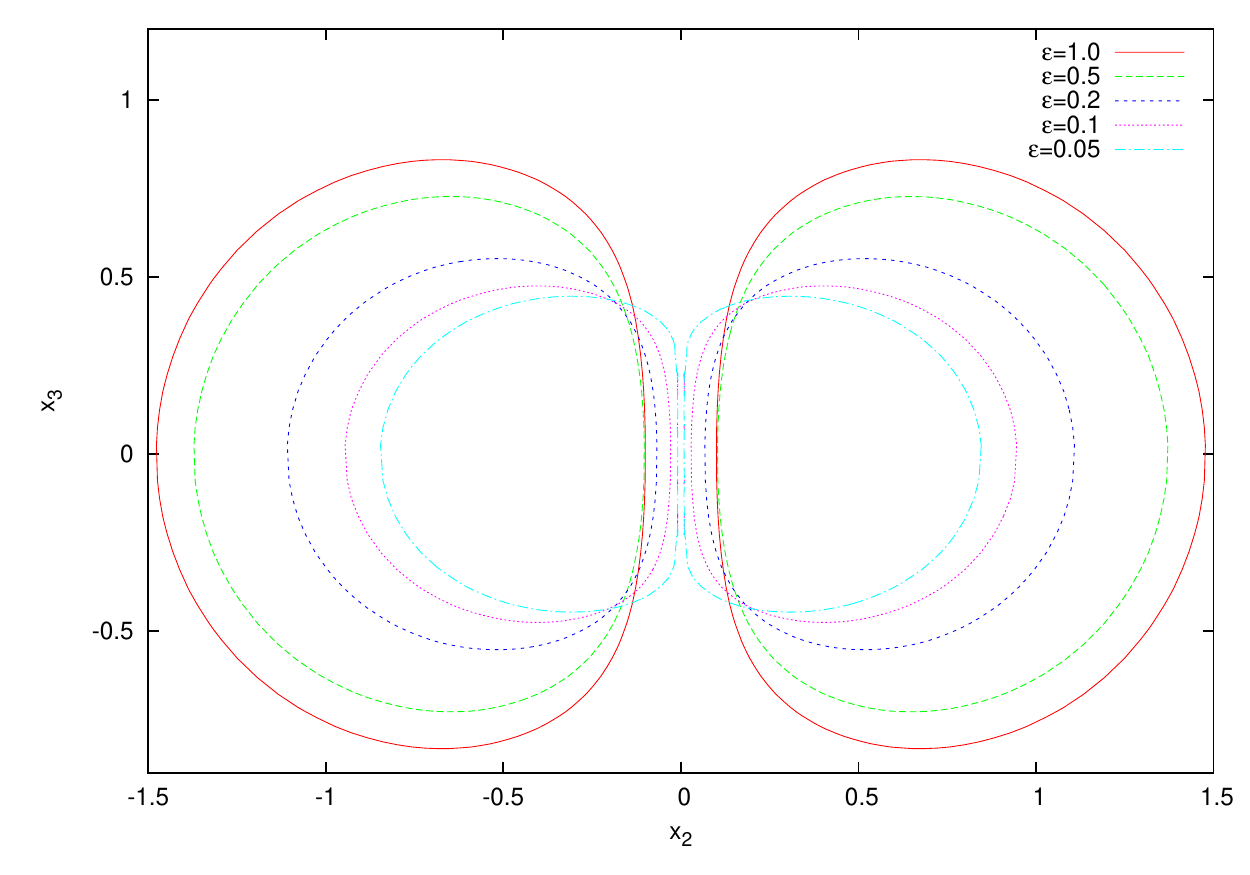}
\end{center}
\caption{Baryon density contours $\mathcal{B}=1.8$ in the plane $x_1=0$ for the 
$B=2$ skyrmion at various values of $\eps$.}
\label{fig:B=2_level_sets}
\end{figure}

As expected,
the $B=1$ minimizer is spherically symmetric, while the $B=2$ minimizer
has deformed toroidal level sets of $\mathcal{B}$. As $\eps\ra 0$, the
hole through the centre of these tori closes up, and the level sets
develop singularities along the symmetry axis, 
see figure \ref{fig:B=2_level_sets}. The reduction in dimension
allows us to use a much finer grid, adapted to resolve the structures 
developing
near the symmetry axis, so we can follow the solutions reliably 
down to much lower $\eps$ than is possible with the fully three-dimensional,
method. The virial quantity $|D(\epsilon)|$ remained less than 1\% for all minimisers that we found ($D$ was not required to be positive in these simulations, as the use of a logarithmic grid for $\rho$ effectively cuts out a neighbourhood of the $x_3$-axis). The difference in energies of the axial and 3D simulation results were $<0.07\%$ for $B=1$ and $<0.18\%$ for $B=2$.

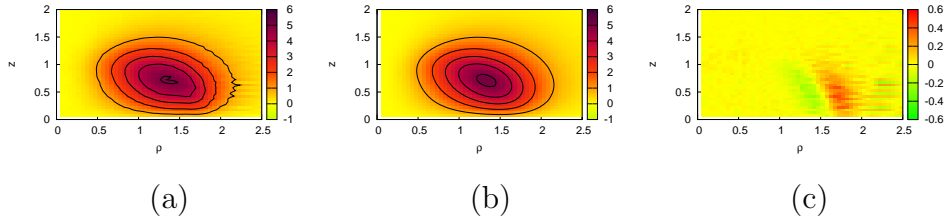
\begin{figure}[htb]
\begin{center}
\begin{picture}(360,100)
 \put(60,0){(a)}
 \put(0,20){\includegraphics[width=4.5cm]{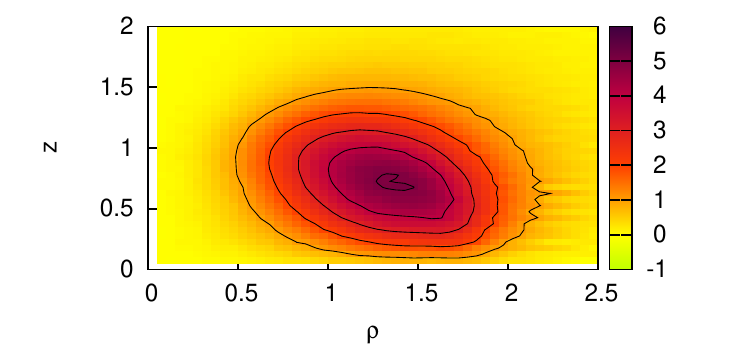}}
 \put(180,0){(b)}
 \put(120,20){\includegraphics[width=4.5cm]{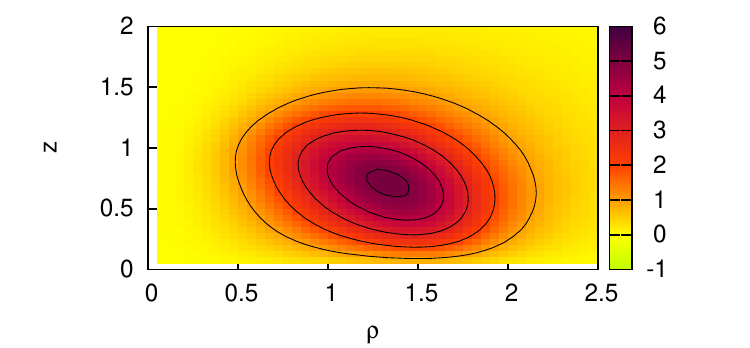}}
 \put(300,0){(c)}
 \put(240,20){\includegraphics[width=4.5cm]{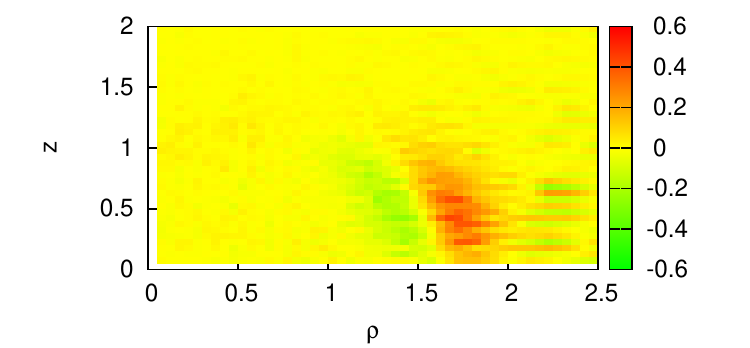}}
\end{picture}
\end{center}
\caption{Density plots for the left (a) and right (b) hand sides of the restricted harmonic equation \eqref{RHeq} evaluated on the soliton with $\epsilon=1$ and $B=2$, and their difference (c).  The plots show that equation \eqref{RHeq} is satisfied to within numerical error.}
\label{fig:reshar}
\end{figure}

Any static solution of the sextic model with $\eps>0$
must satisfy an interesting property
called {\em restricted harmonicity} \cite{spe-rhm}. If $\ph$ is a critical 
point of $E_\eps$, then $E_\eps=E^{(6)}+E^{(0)}+\eps E^{(2)}$ 
is critical with respect to all smooth 
variations of $\ph:\R^3\ra S^3$, and hence, in particular, with respect to
all variations of the form $\ph_t=\ph\circ\psi_t$, where $\psi_t$
is a curve through the identity element in the group of volume preserving
diffeomorphisms of $\R^3$. Now, as already observed, $E^{(0)}$ and
$E^{(6)}$ are invariant under volume preserving diffeomorphisms, so it follows
that $E^{(2)}$ itself must be critical with respect to such variations. This
is precisely the condition for $\ph$ to be restricted harmonic, in the
terminology introduced in \cite{spe-rhm}, where it was shown that
restricted harmonicity is equivalent to the condition that
$\div\ph^*h$ is an exact one-form. Here $h$ denotes the Riemannian metric on
the target space (in this case, $S^3$). Since $H^1(\R^3)=0$, this is
equivalent to the third order nonlinear PDE
\beq
\d(\div\ph^*h)=\cd_k\cd_i(\cd_i\ph\cdot\cd_j\ph)dx_k\wedge dx_j=0.
\label{eq:rhm}
\eeq
For fields within the axially symmetric ansatz (\ref{eq:axan}), this PDE
reduces to
\begin{multline}
\rho^3(\pa_\rho \triangle\nv\cdot\pa_z\nv-\pa_z \triangle\nv\cdot\pa_\rho\nv) +\rho^2(\pa_\rho^2\nv\cdot\pa_z\nv-\pa_\rho\nv\cdot\pa_\rho\pa_z\nv)\\
 = \rho\pa_\rho\nv\cdot\pa_z\nv-2B^2n_1\pa_zn_1,
\label{RHeq}
\end{multline}
in which $\triangle=\pa_r^2+\pa_z^2$.  In figure \ref{fig:reshar} we present evidence that our axially symmetric
$B=2$ numerical solutions are approximately resticted harmonic. The restricted harmonic map
equation is a closed condition on the space of all fields (with respect
to any sensible choice of topology thereon). Since minimizers of $E_\eps$
for $\eps>0$ are restricted harmonic for all $\eps>0$, it follows that, if
the minimizers converge to some limit as $\eps\ra 0$, this limit should
likewise be restricted harmonic. This is another reason to be sceptical
of the axially symmetric BPS skyrmions $\ph_H\circ\psi_B$ used in
\cite{adasanwer,adasanwer2,adanaysanwer,bonmar,bonharmar}: it is known that,
except for $B=1$, none of these solutions are restricted harmonic
\cite{spe-rhm}, and that their failure to be so gets progressively worse
as $B$ increases.

Recall that the underlying motivation for this model is the hope that, for
$\eps$ small and positive, it may support skyrmions with classical
binding energies close to real nuclear physics data. 
So is $\eps=0.2$ small enough? Sadly, no. 
While $E_{0.2}$ supports skyrmions with significantly lower binding energies
than those found in the conventional (massless or massive) 
Skyrme model, they are
still much larger than those found in nature: see figure 
\ref{fig:thatsallfolks} in section \ref{sec:conc}. We can use the 
results of our more
refined axially symmetric simulations to predict roughly how small $\eps$
would need to be to give classical binding energies in the right ballpark.
Real nuclei have binding energies per nucleon of around 1\%, so we
should demand that
\beq
\frac{BE_{min}(1,\eps)-E_{min}(B,\eps)}{BE_{min}(1,\eps)}\approx 0.01.
\eeq
Imposing this for $B=2$ yields $\eps=0.014$, considerably beyond the
reach of our main numerical scheme. (We use the rough figure 1\%
rather than fitting to the binding energy
of the deuteron since the latter is anomalously small.) 
For $\eps$ so small, it is reasonable to hope that $\ph$ will be close to
a BPS skyrmion, that is, a minimizer of $E_0=E^{(6)}+E^{(0)}$.
The question is, which one? Except for $B=1,2$ where enhanced symmetry
allows a reduction in dimension, this question seems to be numerically
intractable. Our results suggest that the minimizers converge as $\eps\ra 0$
to BPS skyrmions of low regularity, which typically have strings of
conical singularities,  and which preserve the property of restricted 
harmonicity.

\section{The lightly bound model}
\label{sec:3}\label{sec:lightly}
\news

The lightly bound model is defined by adding to Skyrme's lagrangian \eqref{Skyrme lagrangian} a term proportional to $\Tr(1-U)^4$.  After fixing units of energy and length, this model depends on two non-trivial parameters.  In suitable units, the associated static energy functional takes the form
\begin{multline}
\label{lightly bound energy}
 E = \int_{\RR^3} \Big[ (1-\alpha)\left( -\frac12\Tr(R_iR_i) + m^2\Tr(1-U) \right) \\ - \frac{1}{16}\Tr([R_i,R_j][R_i,R_j]) + \alpha(\sfrac12\Tr(1-U))^4 \Big] \dd^3 x.
\end{multline}
Here $\alpha\in[0,1]$ and $m:=(2m_\pi\sqrt{1-\alpha}/F_\pi g)\geq0$ are dimensionless parameters.  In writing the energy in this way, we have implicitly chosen $F_\pi/4g\sqrt{1-\alpha}$ as a unit of energy and $2\hbar c\sqrt{1-\alpha}/F_\pi g$ as a unit of length.  The advantage of parametrising the model in this way is that soliton energies and sizes remain roughly constant as $\alpha$ is varied, thus enabling simulations with a range of values of $\alpha$ to be performed on the same grid.  In the limiting case $\alpha=1$ the above model has a topological energy bound $E\geq 8\pi^2|B|$ which can be saturated only in the $|B|=1$ sector \cite{harland}. Solitons with $|B|>1$ are therefore unstable to fission.  The other extreme case $\alpha=0$ is the standard massive Skyrme model, in which binding energies are too large.

Numerical approximations to minima of the lightly bound energy \eqref{lightly bound energy} with $B$ between one and eight have been determined for a range of values of $\alpha$ and for $m=1$.  Fixing the value of $m$ ensures that the Compton wavelength of the pion remains comparable to the size of a nucleus, as nuclear radii do not depend strongly on the value of $\alpha$.

To simulate the model for a given $\alpha$ numerically we used the same numerical scheme as in section \ref{sec:sextic} on a cubic lattice of size $N^3$ with $N=301$ and lattice spacing $h=0.05$. We obtained a lattice approximant to the energy for a given value of $\alpha$, $E_\alpha$ for one or more (perhaps only local) energy minima in each charge sector. A wide range of initial conditions were used to assist in finding the global energy minimimiser for each baryon number considered. The accuracy checks, already described, were applied.  
Firstly, the baryon number was calculated using the formula \eqref{baryon number} and evaluated using the same scheme. The value obtained was correct to within 0.001\% of the integer value.  Secondly, the virial constraint $D(\alpha)$, similar to (\ref{eq:virial}), was evaluated on energy minima (in which $E^{(a)}$ denotes the component of the energy involving $a$ derivatives):
\beq
D(\alpha):=\frac{1}{E}(E^{(4)}-E^{(2)}-3E^{(0)})=0.
\label{eq:virial_LB}
\eeq 
When evaluated on our numerically-determined minima this quantity was at worst 0.01\%.  

For $B$ between one and four the full range of $\alpha$ was explored.  Two sets of simulations were performed: one in which the minima were initially determined for $\alpha=0.95$ and tracked as $\alpha$ decreased to 0 in steps of 0.05, and one in which the minima were initially determined for $\alpha=0$ and tracked as $\alpha$ increased to 0.95 in steps of 0.05.  Both sets of simulations gave similar results for $\alpha$ close to 0 and 0.95. For small $\alpha$ the minima were qualitatively similar those of the Skyrme model and in particular shared their symmetry groups. For $\alpha$ near 0.95 energy densities of minima were localised at $B$ points, which we interpret as the locations of the $B$ nucleons within the nucleus. The transitions between the two types of soliton occured at different values of $\alpha$ according to whether $\alpha$ was increasing or decreasing.  For $\alpha$ increasing the transitions occured when $\alpha\in[0.4,0.5]$, while for $\alpha$ decreasing they occured when $\alpha\in[0.35,0.4]$.

\begin{table}[htb]
\begin{center}
\begin{tabular}{|l|l|l|l|l|}
\hline
Configuration&	$E/8\pi^2$ &	$\sqrt{\langle r^2\rangle_{I=0}}$&	BEPN (MeV)&	$\sqrt{\langle r^2\rangle_{I=0}}$ (fm) \\
\hline
1	&	1.1082	&	1.2457	&	0.00	&	0.8104	\\
2	&	2.2119	&	2.2261	&	1.84	&	1.4481	\\
3	&	3.3129	&	2.7627	&	3.26	&	1.7971	\\
4	&	4.4111	&	2.9310	&	4.56	&	1.9066	\\
5	&	5.5132	&	3.4593	&	4.66	&	2.2503	\\
6a	&	6.6128	&	3.4426	&	5.09	&	2.4190	\\
6b	&	6.6130	&	3.4426	&	5.07	&	2.2394	\\
7a	&	7.7120	&	3.5350	&	5.45	&	2.2995	\\
7b	&	7.7127	&	3.8888	&	5.36	&	2.5297	\\
7c	&	7.7140	&	4.1668	&	5.21	&	2.7105	\\
8a	&	8.8090	&	4.0367	&	5.95	&	2.6259	\\
8b	&	8.8094	&	3.9610	&	5.91	&	2.5766	\\
8c	&	8.8124	&	4.1150	&	5.59	&	2.6768	\\
8d	&	8.8130	&	4.1431	&	5.52	&	2.6951	\\
8e	&	8.8130	&	4.1430	&	5.52	&	2.6950	\\
\hline
\end{tabular}
\caption{Energies, isoscalar charge radii, and binding energies per nucleon of skyrmions in the lightly bound model with $\alpha=0.95$.}
\label{LB data}
\end{center}
\end{table}

\begin{figure}[htb]
\begin{center}
\includegraphics[width=13cm]{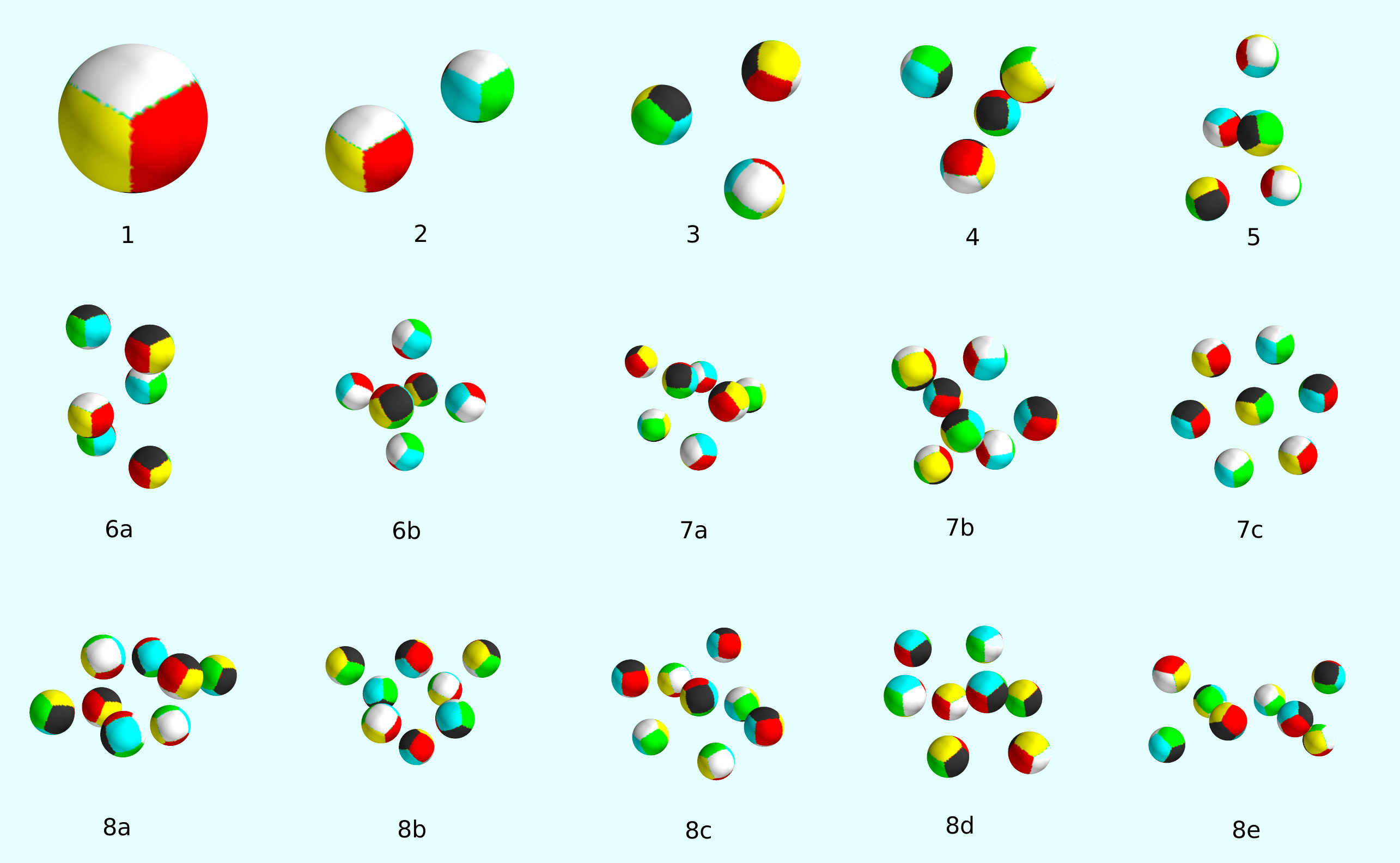}
\end{center}
\caption{Baryon density isosurfaces of skyrmions in the lightly bound model.}
\label{LB densities}
\end{figure}

\begin{figure}[htb]
\begin{center}
\includegraphics[width=13cm]{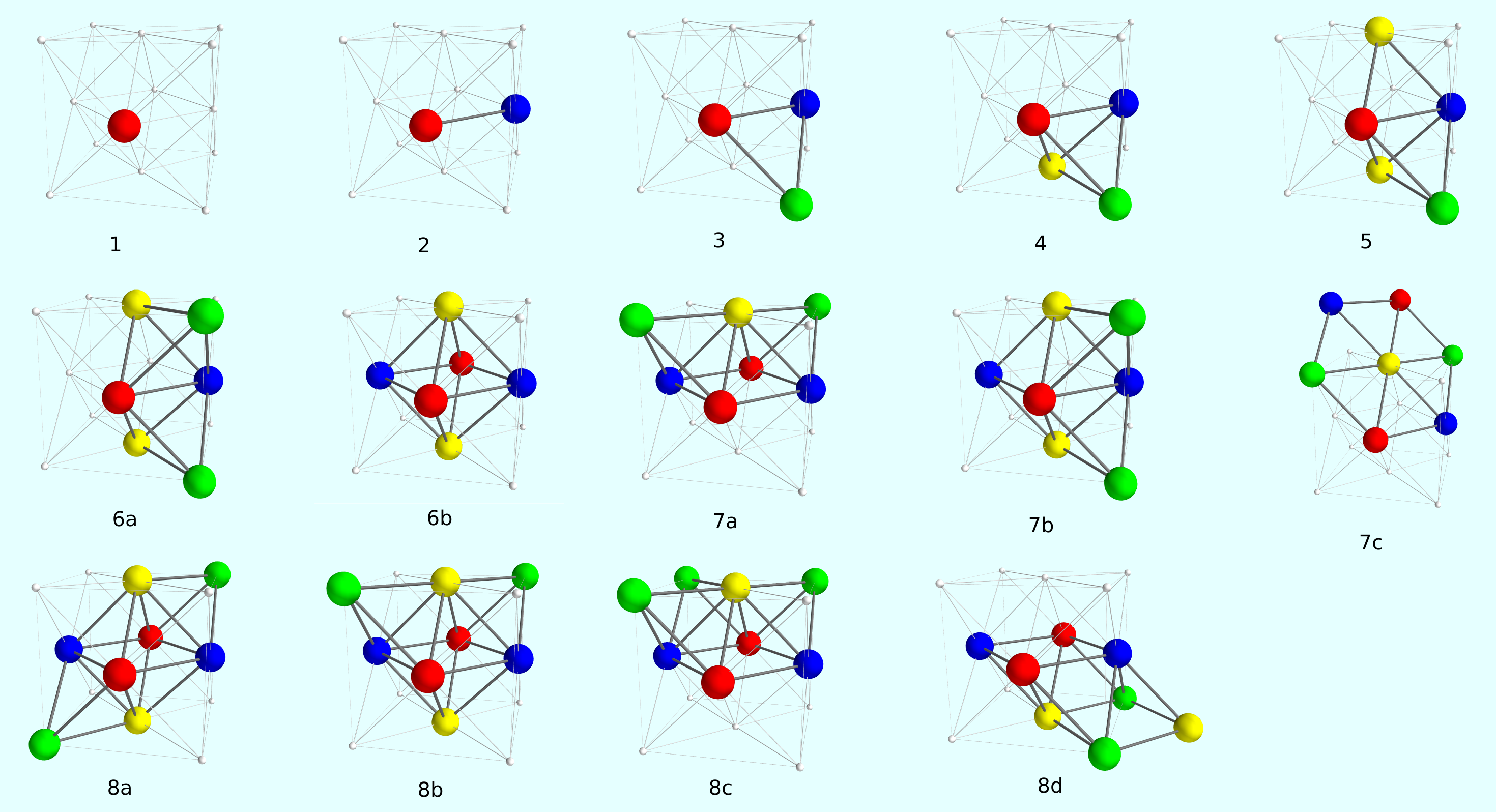}
\end{center}
\caption{Subsets of the face-centred-cubic lattice corresponding to the isosurfaces in figure \ref{LB densities}.}
\label{LB lattice}
\end{figure}

Energy minima were examined more carefully for $\alpha\in[0.9,1.0]$ and $B$ between one and eight.  This range of $\alpha$ is the one in which binding energies are comparable to experimental values.  We found that $\alpha=0.95$ gave binding energies of appropriate magnitude relative to the proton mass; baryon density isosurfaces for the minima that we found are displayed in figure \ref{LB densities}.  Note that for $B\geq 6$ several local minima of the energy have been found.  The isosurfaces have been coloured to indicate the orientation of the pion fields on the surface, as described in the previous section.  The energies of these configurations are recorded in table \ref{LB data}.  Also recorded in this table are their charge radii $\sqrt{\langle r^2\rangle_{I=0}}$, defined by the formulae
\begin{equation}
 \langle r^2\rangle_{I=0} = \frac{1}{2\pi^2B}\int_{\RR^3} \|\mathbf{x}-\mathbf{x}_0\|^2\, \mathcal{B}(\mathbf{x})\,\dd^3 x,\quad \mathbf{x_0}=\frac{1}{2\pi^2B}\int_{\RR^3} \mathbf{x}\,\mathcal{B}(\mathbf{x})\,\dd^3 x.
\end{equation}

The most striking feature of the baryon density isosurfaces is that they display a clear particle-like structure.  This is in marked contrast with the traditional Skyrme model, in which individual nucleons are not discernible within a nucleus. The value of $0.1\max\{\mathcal{B}\}$ used to create the isosurfaces in figure \ref{LB densities} is significantly lower than those commonly used in the Skyrme model; if instead a larger value of $\mathcal{B}$ is used the particle-like structure is even more pronounced.  The points at which $\mathcal{B}$ attains local maxima are close to the $B$ points at which the field $U$ takes the value $-1$.  Unlike in the usual Skyrme model these preimage points are non-degenerate.  The likely cause of this is the choice of potential in our model, which strongly disfavours $U$ being close to $-1$.  The standard potential $\Tr(1-U)$ has a similar but weaker effect: with this potential $-1$ has two preimage points with four-fold degeneracy in the $B=8$ soliton, whereas without it $-1$ has a single preimage point with eight-fold degeneracy.

Another remarkable feature evident in figure \ref{LB densities} is that, with one exception, all configurations resemble subsets of a face-centred cubic lattice.  Within each of these lattice-like configurations, no more than four distinct orientations for the constituent particles can be observed (corresponding to whether the particle is on a vertex of a cube, a horizontal face, or either of the two orientations for vertical faces).  The precise subsets of the lattice are displayed in figure \ref{LB lattice}, with the four orientations distinguished by four colours.  The only soliton to which these comments do not apply is configuration 8e.  This configuration resembles a pair of adjacent tetrahedra and cannot be realised as a subset of the face-centred cubic lattice; its eight constituent particles all have different orientations.  

\begin{figure}[htb]
\begin{center}
\includegraphics{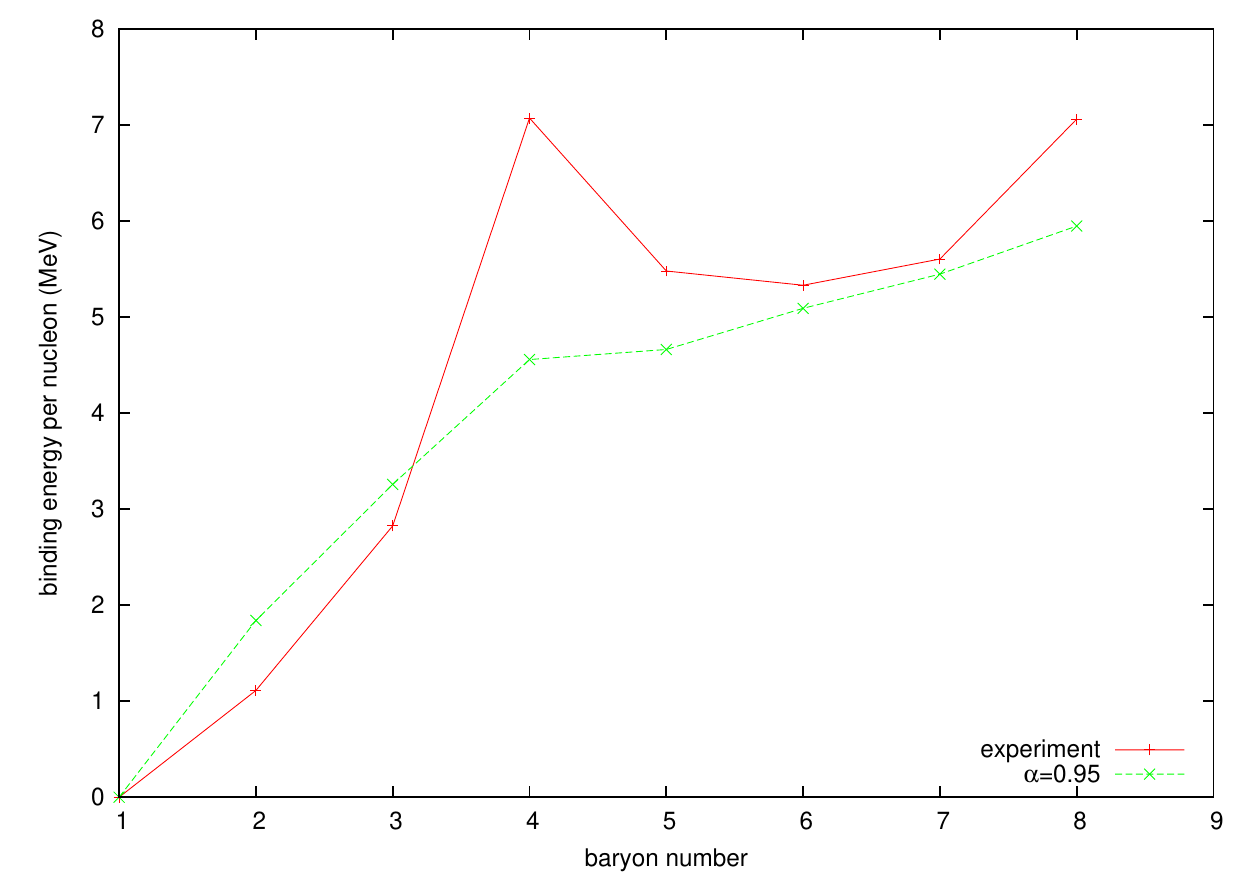}
\end{center}
\caption{Binding energy per nucleon as a function of baryon number in the lightly bound model and from experimental nuclear physics.}
\label{LB binding energies}
\end{figure}

\begin{figure}[htb]
\begin{center}
\includegraphics{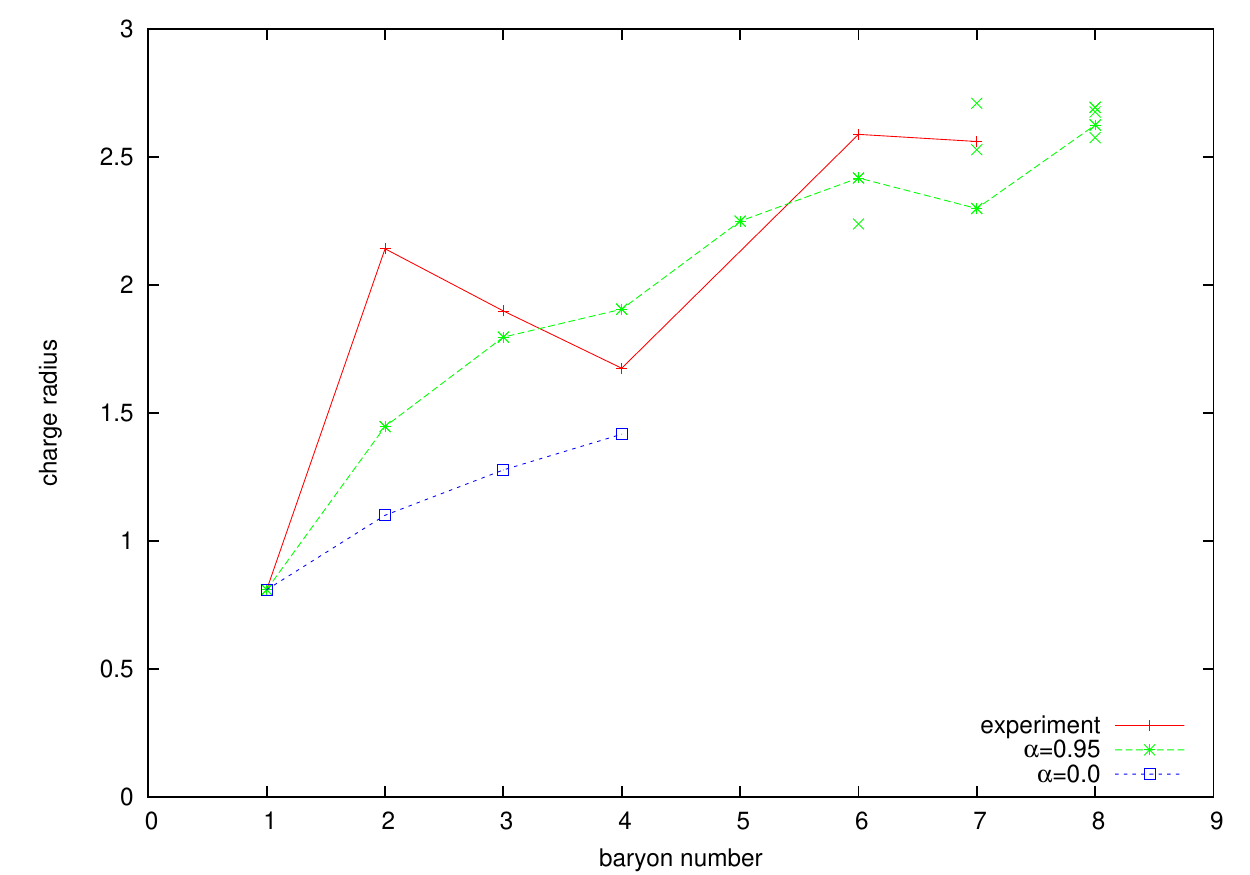}
\end{center}
\caption{Isoscalar charge radii in the lightly bound model ($\alpha=0.95$), the standard Skyrme model ($\alpha=0$), and from experimental nuclear physics.  The $\alpha=0.95$ curve passes through the points corresponding to global energy minima.}
\label{LB charge radii}
\end{figure}

In order to make a direct comparison between the lightly bound model and experimental nuclear physics we have calibrated our units of energy and length.  The energy units were chosen so that the one-soliton had energy equal to the proton mass, 938.27MeV.  The length units were chosen by fitting the electric charge radii of the proton and neutron.  If the proton and neutron are modelled as rigidly isospinning bodies centred at the origin then their charge radii are given by
\begin{equation}
 \langle r^2\rangle_p = \int_{\RR^3} r^2 \left(\sfrac{1}{4\pi^2} \mathcal{B}+ \mathcal{I}_3\right)\dd^3x,\quad \langle r^2\rangle_n = \int_{\RR^3} r^2 \left(\sfrac{1}{4\pi^2} \mathcal{B}- \mathcal{I}_3\right)\dd^3x,
\end{equation}
where $\mathcal{I}_3$ denotes the isospin charge density, normalised such that $\int_{\RR^3} \mathcal{I}_3\dd^3 x=\frac12$.  We therefore identify
\begin{equation}
\label{B=1 charge radius}
 \sqrt{\langle r^2\rangle_{I=0}} = \sqrt{\langle r^2 \rangle_p + \langle r^2\rangle_n} = \sqrt{0.8783^2-0.1149}\,\mathrm{fm}=0.8103\,\mathrm{fm}.
\end{equation}

These normalisation conventions force us to identify
\begin{equation}
 \frac{F_\pi}{4g\sqrt{1-\alpha}} = 10.72 \,\mathrm{MeV},\quad \frac{2\hbar c\sqrt{1-\alpha}}{F_\pi g} = 0.6505 \,\mathrm{fm}.
\end{equation}
Hence the following values are chosen for the parameters in the Skyrme lagrangian \eqref{Skyrme lagrangian}:
\begin{equation}
 F_\pi = 29.1\,\mathrm{MeV},\quad m_\pi = 303\,\mathrm{MeV},\quad g = 3.76.
\end{equation}
With these units the pion decay constant is significantly below its experimental value of 186MeV, and the pion mass is significantly above its experimental value of 138MeV.  Most calibrations of the standard Skyrme model result in abnormally low values of the pion decay constant.  Although this problem seems to be exacerbated in the lightly bound model, we expect for reasons to be explained below that the value of $F_\pi$ would increase in a quantised treatment of the model.  A better value for the pion mass could be obtained by reducing the value of the parameter $m$ in the model, but we have not explored this possibility.  Reducing the value of $m$ would probably reduce soliton energies, and thereby also increase the value of $F_\pi$.

The binding energy per nucleon is plotted alongside experimental values in figure \ref{LB binding energies}.  The theoretical binding energy per nucleon is calculated from the formula $E_B/B-E_1$, in which $E_B$ represents the global minimimum of the energy in the charge $B$ sector.  The corresponding experimental values are the binding energies for the lightest isotope at each mass number $B$.  The theoretical curve closely follows the experimental curve for most values of $B$ in the range considered, except that the anomalies at $B=4$ and 8 are less pronounced in the theoretical curve than in the experimental curve.  This graph represents a substantial improvement on the standard Skyrme model (see figure \ref{fig:thatsallfolks} for a comparison).

The skyrmion isoscalar charge radii are plotted alongside experimental values in figure \ref{LB charge radii}.  For $B$ even the experimental values shown are the charge radii of isotopes with mass number $B$ and isospin 0.  For $B$ odd they are calculated as weighted averages of the states with isospin $I=\pm 1/2$ via the formula $\langle r^2\rangle_{I=0} = [(B+1)\langle r^2\rangle_{I=1/2}+(B-1)\langle r^2\rangle_{I=-1/2}]/2B$.  Like equation \eqref{B=1 charge radius}, this formula is motivated by identifying the $I=\pm1/2$ states with solitons isorotating rigidly in opposite directions.

Figure \ref{LB charge radii} shows that the theoretical and experimental charge radii are comparable.  The most surprising feature of the experimental curve is the abnormally large radius at $B=2$ in comparison to $B=1$.  The lightly bound model does not fully capture this feature, but approximates it better than the standard Skyrme model, which is plotted alongside for comparison.


\section{Concluding remarks}
\label{sec:4}\label{sec:conc}

\begin{figure}[htb]
\begin{center}
\includegraphics[width=13.5cm]{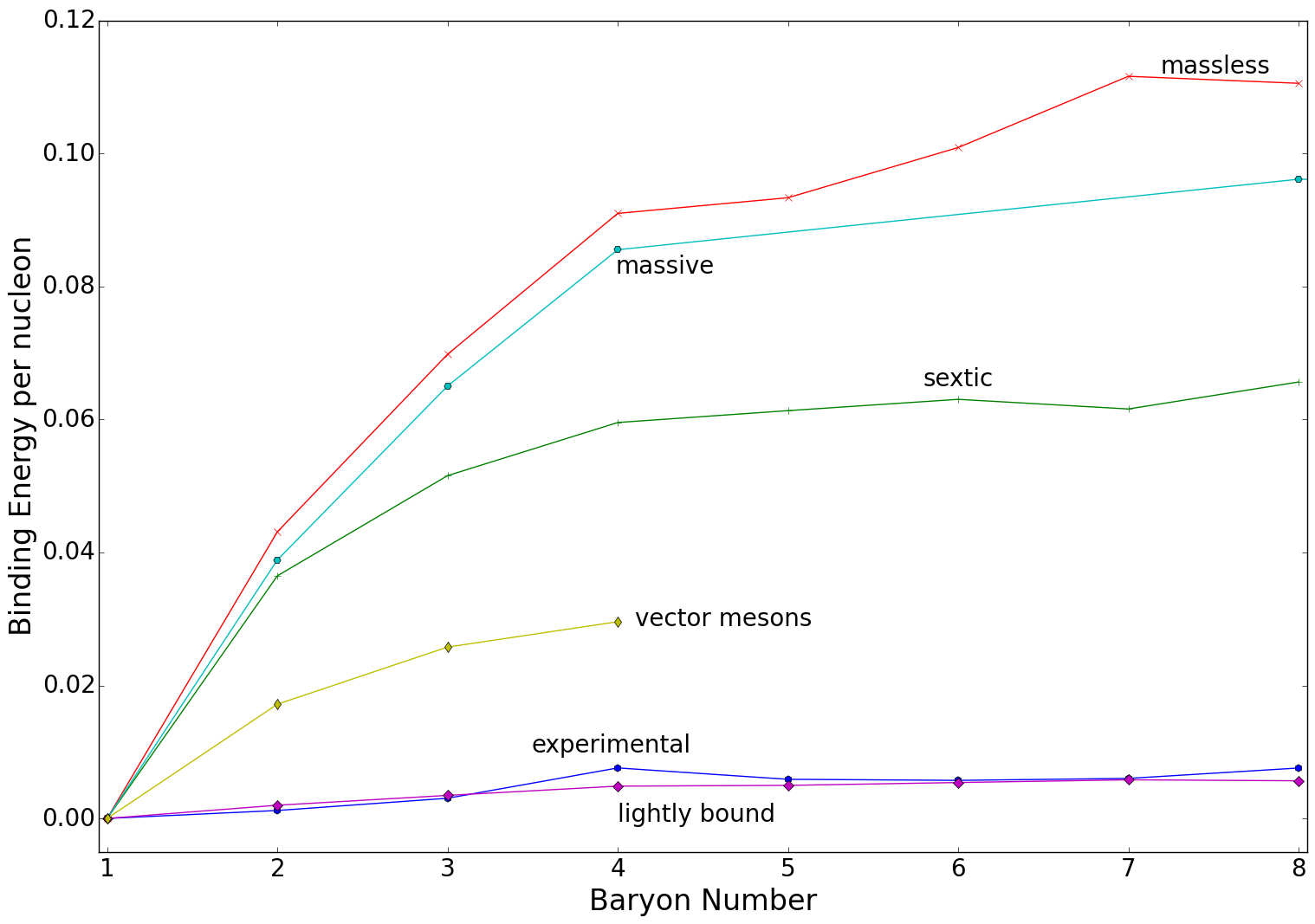}
\end{center}
\caption{Binding energy per nucleon for $B=2,\ldots,8$ in various variants of the Skyrme model: the \textit{massless} Skyrme model with $m=0$, the \textit{massive} Skyrme model with $m=1$, the \textit{sextic} Skyrme model with $\eps=0.2$, holographic Skyrme model with \textit{vector mesons}, the \textit{lightly bound} Skyrme model with $\alpha=0.95$ and \textit{experimental} data. Markers indicate values of available data.}
\label{fig:thatsallfolks}
\end{figure}

We have investigated the problem of obtaining low binding energies in two variants of the Skyrme model.  Our results are summarised in figure \ref{fig:thatsallfolks}, alongside results obtained from a third ``holographic'' variant explored in \cite{sut-holographic}, the standard Skyrme model, and the experimental binding energy curve.  The sextic and holographic models are both capable, in principle, of producing skyrmions with very low binding energies, but in practice they are yet to yield realistic binding energies in numerical simulations. For the sextic model, the problem is that the skyrmions start to develop singularities, and become numerically inaccessible, before the regime of very low binding energy is reached. For the holographic model, which is a Skyrme model coupled to an infinite tower of vector mesons, only the lowest truncation level, including just the lightest vector meson and the lightest axial vector meson, has been treated numerically. One could, in principle, include more mesons from the tower, and this would presumably reduce the binding energies
further. However the resulting model would be formidably complicated, and hence difficult to solve numerically in practice.  In contrast, the lightly bound model looks very promising: using standard numerical algorithms, binding energies have been obtained which are comparable with experimental values. This model clearly deserves further attention.

The solitons obtained in the lightly bound model differ markedly from those obtained in the standard Skyrme model, in that they resemble a lattice of nucleons.  This result suggests that the problem of finding solitons with higher baryon number could be simplified by first investigating a point particle model, along the lines of \cite{salsut}.

Our analysis of the lightly bound model has been purely classical, but in the future this model should be quantised semi-classically.  Including quantum corrections would no doubt modify some of our results.  Most notably, the masses of quantised states would be higher than the classical values quoted here, with the one-soliton affected more severely than others.  Therefore binding energies are likely to be reduced in a semi-classical treatment, and the value chosen for $\alpha$ would need to be reduced to compensate for this.  Reducing the value of $\alpha$ would also lead to a more realistic value for the pion decay constant.

Our attempts to obtain realistic binding energies within the sextic model have been hampered by numerical difficulties associated with the formation of singularities in the field.  Thus a different numerical scheme is called for if any progress is to be made with this model.  Despite these difficulties, we have found clear evidence that solitons in this model do not resemble the ansatz proposed in \cite{adasanwer,adasanwer2,adanaysanwer,bonmar,bonharmar}.  In this light it would be worthwhile therefore to revisit some of the results obtained therein.

\subsection*{Acknowledgements}

This work was supported by EPSRC through the grant
 ``Geometry, Holography and Skyrmions.''
We acknowledge valuable conversations with Christoph Adam,
Andrzej Wereszczynski, Paul Sutcliffe, and Nick Manton.
The majority of simulations were performed using computer code 
originally developed in collaboration with Juha J\"aykk\"a and utilised the ARC High Performance Computing facilities at the University of Leeds.

\bibliographystyle{./h-elsevier}
\bibliography{./bibliography}


\appendix
\section{Appendix: energies in the sextic model}
\label{sec:6}\label{sec:apndx}


\begin{table}[htb]
\begin{center}
\begin{tabular}{|l|l|l|l|l|l|}
\hline
Configuration &  \multicolumn{2}{c|}{$\epsilon=1$} & & \multicolumn{2}{c|}{$\epsilon=0.2$} \\
\hline
\hline
&	$E_1$ &	$B$ &	&	$E_{0.2}$ &	$B$ \\
\hline
1	&	62.7543	&	0.999991	& &	29.4172	&	0.999995	\\
2	&	116.696	&	1.999981	& &	56.6878	&	1.999972	\\
3	&	169.146	&	2.999970	& &	83.6968	&	2.999764	\\
4	&	220.424	&	3.999964	& &	110.659	&	3.999790	\\
5	&	274.815	&	4.999958	& &	138.065	&	4.999874	\\
6	&	327.445	&	5.999950	& &	165.374	&	5.999737	\\
7	&	379.956	&	6.999939	& &	193.238	&	6.999792	\\
8a	&	433.719	&	7.999931	& &	220.633	&	7.999514	\\
8b	&	435.214	&	7.999934	& &	219.884	&	7.999600	\\
\hline
\end{tabular}
\caption{Numerical values of energy and baryon number for skyrmions in figure \ref{fig:leaf}.  $E_\epsilon$ is given by (\ref{eq:nearbps}).}
\label{tb:bpsdata}
\end{center}
\end{table}

\end{document}